\documentclass[3p,times,procedia]{elsarticle}
\usepackage{nupha_ecrc}
\usepackage{subcaption}
\usepackage{array}
\usepackage{geometry}
\usepackage{wrapfig}
\usepackage{lipsum}
\usepackage[font=footnotesize,skip=0pt]{caption}
\usepackage{sidecap}

\volume{00}

\firstpage{1}

\journalname{Nuclear Physics A}

\runauth{}

\jid{nupha}

\jnltitlelogo{Nuclear Physics A}

\usepackage{amssymb}

\usepackage[figuresright]{rotating}

\newcommand{\piz}{{\mathbf{\pi^{0}}}}

\newcommand{\pp}{{pp\,}}

\newcommand{\pbpb}{Pb-Pb\,}

\renewcommand{\aa}{{\rm AA}}

\newcommand{\pt}{{$p_{\rm T}$}}

\newcommand{\raa}{{R_\aa}}

\newcommand{\GeVc}{GeV/$c$}

\begin{document}

\begin{frontmatter}



\dochead{}

\title{Measurement of neutral mesons in pp and Pb-Pb collisions at mid-rapidity with ALICE}


\author{Astrid Morreale on behalf of the ALICE collaboration}

\address{Laboratoire de Physique Subatomique et des Technologies Associ\'{e}es SUBATECH.\\
4 Rue Alfred Kastler, 44300 Nantes, France}

\begin{abstract}
One of the key signatures of the Quark-Gluon Plasma (QGP), is the modification of hadron transverse momentum differential cross-sections in heavy-ion collisions (HIC) as compared to proton-proton (pp) collisions.  Suppression of hadron production at high transverse momenta (\pt)~in HIC has been explained by the energy loss of the partons produced in the hard scattering processes which traverse the deconfined quantum chromodynamic (QCD) matter. The dependence of the observed suppression on the \pt~ of the measured hadron towards higher \pt~  is an important input for the theoretical understanding of jet quenching effects in the QGP and the nature of the energy loss. The ALICE experiment at the Large Hadron Collider (LHC) performs measurements of neutral meson inclusive spectra at mid-rapidity in a wide \pt~ range in $pp$, $p$-Pb and Pb-Pb collisions. Neutral mesons ($\pi^{0}$, $\eta$, $\omega$) are reconstructed via complementary methods, using the ALICE electromagnetic calorimeters, PHOS and EMCal, and by the central tracking system, identifying photons converted into $e^+e^-$ pairs in the material of the inner barrel detectors (TPC and ITS).  
In this presentation, an overview of  current $\pi^{0}$ and $\eta$ measurements  by the ALICE experiment in HIC and pp collisions will be given. 

\end{abstract}
\end{frontmatter}


\section{Light mesons as probes of the QGP}\label{intro}
\vspace*{-0.2cm}
Light mesons offer an opportunity to study and quantify QCD processes with a variety of observables.
In pp collisions, hadrons are produced from parton fragmentation in the QCD vacuum. The study of hadron production in pp and and heavy-ion collisions (HIC) gives information of parton fragmentation and  parton distribution functions (pdf) as well as non-linear recombination effects~\cite{Lappi:2013zma}. Within the Quark Gluon Plasma (QGP) and in central heavy ion collisions, scattered partons  interact strongly with the medium leading to modifications of parton fragmentation. These effects can be observed via inclusive spectra modifications and quantified further with particle ratios comparisons in HIC and the corresponding measurements in pp collisions. 
Of the many observables accessible in HIC which probe the QGP, measurements of $\piz$ and  $\eta$  production in different colliding systems are of particular interest. The physics motivation behind these measurements are different, depending on the \pt~range of the measured neutral meson. 

At low \pt~(\pt~$\lesssim 3-5~$GeV/$c$), mesons give insight about bulk properties and collective effects.  
At collider energies, gluon fragmentation contributes significantly to $\pi^{0}$ and $\eta$ production. Due to the different color factors of quark-gluon vertex and gluon-gluon vertex it is predicted that
gluons will  suffer a larger energy loss in the medium than quarks. This difference in parton energy loss, together with their relative contribution to $\pi^{0}$ and $\eta$ production may give rise to differences in suppression patterns between $\piz$~ and $\eta$~\cite{Dai:2015dxa, Guo:2000nz}. 
At high \pt~(\pt~$\gtrsim 5~$GeV/$c$), hadron production results from the hadronization of partons created in initial hard scattering. As partons interact strongly with the QGP they will lose energy via radiation and collisions.
We can measure high \pt~hadrons, hoping that they carry a large fraction of their parent parton's energy and 
interpret the suppression of hadron production at high \pt~in HIC as the energy loss of the scattered parton in the QGP.  
\vspace*{-0.5cm}
\section{ALICE's neutral meson reconstruction}\label{reco}
\vspace*{-0.2cm}
The ALICE detector~\cite{ALICE} measures photons with three fully complementary methods: photon conversion method (PCM) and electromagnetic shower calorimetry using two detectors: PHOS~\cite{PHOS} and EMCal~\cite{Abelev:2014ffa}. 
The PCM method measures photons and meson yields by reconstructing $e^{+}e^{-}$ pairs proceeding from photon conversions in the material of ALICE inner detectors. PCM measurements use the ALICE inner tracking system (ITS~\cite{ALICE_ITS}) and the time projection chamber (TPC~\cite{ALICE_TPC}). The electromagnetic calorimeters PHOS and EMCal are based on energy measurements via total absorption of particles.\\ 
\setlength{\intextsep}{-14pt}%
\setlength{\columnsep}{8pt}%
\begin{wrapfigure}[32]{l}{0.42\textwidth}
\begin{center}
\includegraphics[width=0.42\textwidth]{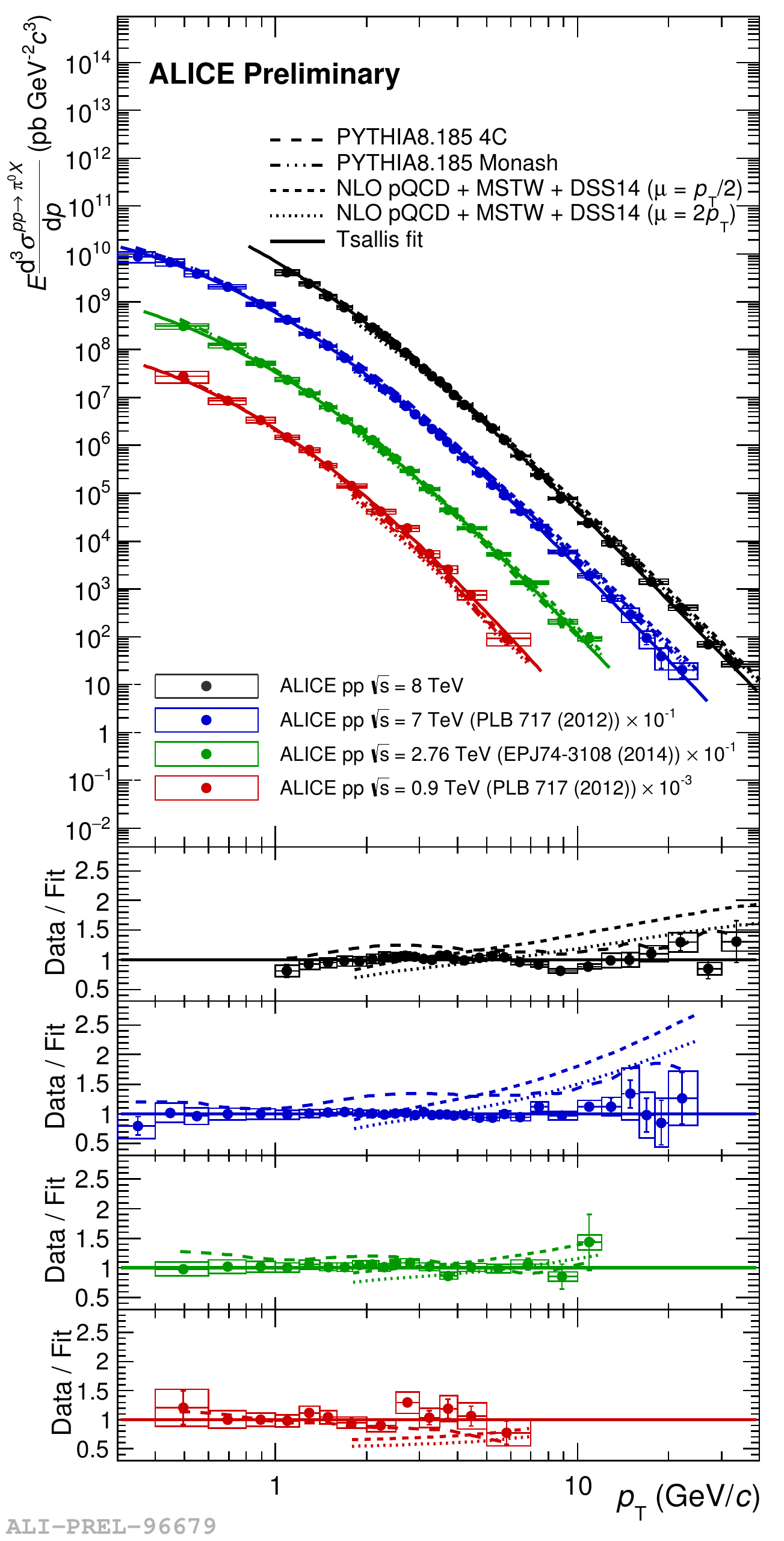}
\captionof{figure}{\footnotesize $\piz$ production in pp collisions measured by ALICE in four center-of-mass energies and compared to NLO pQCD calculations.}\label{fig:ppyields}
\end{center}
\end{wrapfigure}
\vspace{-0cm}
The conversion point can be reconstructed with a $z$ and $\phi$ resolution of 1.5~cm and 7~mrad respectively. 
While photon conversion probability $(\sim8.5\%)$ is  small it is compensated by the measurement's wide acceptance: full azimuthal acceptance and 1.8 units of pseudo rapidity coverage.
The PHOS detector is a high granularity and high energy resolution calorimeter~($\sigma_{E(\rm GeV)}/E=0.01/E\oplus0.04/\sqrt{E}\oplus0.01$) composed of PbWO$_4$ crystals 
 at a 4.6 m radius from ALICE's interaction point (IP). It covers a pseudorapidity range of 0.26 units and has an azimuthal coverage of $60^\circ$.  
The EMCal detector~\cite{Cortese:2008zza} is a modular sampling calorimeter, composed of 77 alternating layers of 1.4 mm lead  and  1.7 mm scintillator. Its energy resolution is $\sigma_{E(\rm GeV)}/{E}=0.05/E\oplus 0.11/\sqrt{E}\oplus0.02$. It is 
located at a 4.28 m radius from ALICE's IP, covers a pseudorapidity range of 1.4 units and has $100^\circ$ of azimuthal coverage.

\vspace*{-0.5cm}
\section{$\piz$ invariant yields in pp collisions at $\sqrt{s} =$ 0.9, 2.76, 7 and 8 TeV}
\vspace*{-0.2cm}
ALICE has measured $\piz$ invariant yields in pp collisions at four different center of mass (CM) energies  $\sqrt{s}=$ 0.9, 2.76, 7~\cite{Abelev:2012cn} and most recently at 8  TeV (Fig.~\ref{fig:ppyields}). 

The pp spectra presented in Fig.~\ref{fig:ppyields} are described by PYTHIA as well as a variety of functions including  Tsallis, Hagedorn and a power law. The measurements indicate a power law dependence at high \pt~with  a measured power value of $n=6.0\pm0.1$ at $\sqrt{s}=$~2.76~TeV, a value  lower than what has been observed with measurements performed at lower CM energies~\cite{Adare:2013esx}. The measurements are also compared to next-to-leading order pQCD (NLO pQCD) which use MSTW pdfs, and DSS14 fragmentation functions (FF)~\cite{deFlorian:2014xna}. These comparisons as well as the ratio are illustrated in Fig.~\ref{fig:ppyields}. 
The predictions with MSTW pdf and DSS14 FF describe the magnitude better than previous pQCD calculations~\cite{Abelev:2012cn}. However an increasing discrepancy between pQCD and the measurements is observed with increasing CM energy and \pt.  

\vspace*{-0.5cm}
\section{$\piz$ in \pbpb and the nuclear modification factor $\raa$}
\vspace*{-0.2cm}


 $\piz$ invariant yields have been previously measured by ALICE in Pb-Pb collisions and in six centrality classes: 0-5$\%$, 5-10$\%$, 10-20$\%$, 20-40$\%$, 40-60$\%$ and 60-80$\%$~\cite{Abelev:2014ypa}.
Nuclear effects in HIC are quantified using the nuclear modification factor $R_{\aa}$ defined as: $R_{\aa}(p_{\rm T})=\frac{1}{N_{\rm coll}}\frac{dN_{\aa}/dp_{\rm T}}{dN_{\rm pp}/dp_{\rm T}}$.
 $R_{\aa}$  compares the yields (dN/d\pt) measured by ALICE ~\cite{Abelev:2014ypa} to  production in pp collisions from Fig.~\ref{fig:ppyields} at the same $\sqrt{s}$. This ratio is scaled by the number of binary nucleon-nucleon collisions (${N_{\rm coll}}$) which are obtained from Glauber Monte Carlo simulations~\cite{Glauber, Abelev:2012hxa}. The measurement of ${R_{\aa}}$ is interpreted as an interplay between initial (Cronin, nuclear shadowing) and final state effects such as jet quenching. These effects would need the corresponding measurement in proton-heavy ion (p-A) collisions for proper disentanglement and interpretation. 
\begin{wrapfigure}[16]{l}{0.4\textwidth}
\begin{center}
\includegraphics[width=0.4\textwidth]{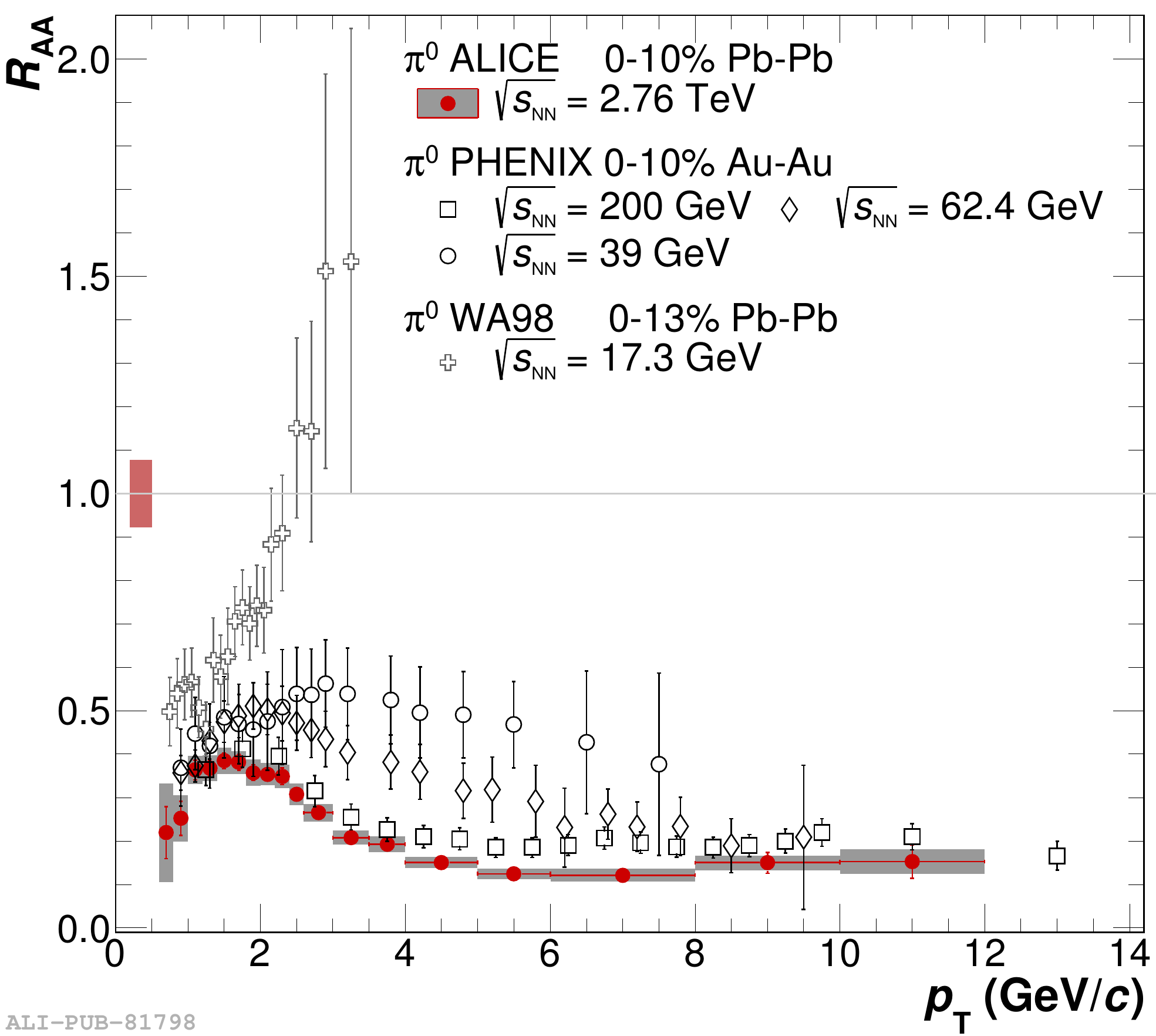}
\end{center}
\captionof{figure}{\footnotesize ${\piz}$ ${R_{\aa}}$, in {\pbpb} collisions at ${\sqrt{s_{NN}}} =$ 2.76 TeV for the $0-10\%$ class in comparison to corresponding measurements at lower energies~\cite{Abelev:2014ypa}.}\label{RAA_0010_Paper}
\end{wrapfigure}
ALICE has measured ${\piz}$ $R_{\aa}$ in six centrality classes and has found 
a large ${\piz}$ suppression in central \pbpb collisions as it can be seen in Fig.~\ref{RAA_0010_Paper}. In addition, ALICE data when compared to the world measurements at different $\sqrt{s}$~\cite{Abelev:2014ypa, PhysRevLett.109.152301, Adare:2008qa,Adare:2012uk, PhysRevLett.100.242301}, indicates an energy dependence with the ALICE's highest $\sqrt{s}$ data points exhibiting the highest suppression (Fig.~\ref{RAA_0010_Paper}).
\vspace*{-0.4cm} 
\section{New results:~$\piz$ and $\eta$ in Pb-Pb collisions}
\begin{figure}
\begin{subfigure}{0.35\textwidth}
\includegraphics[width=5.0cm]{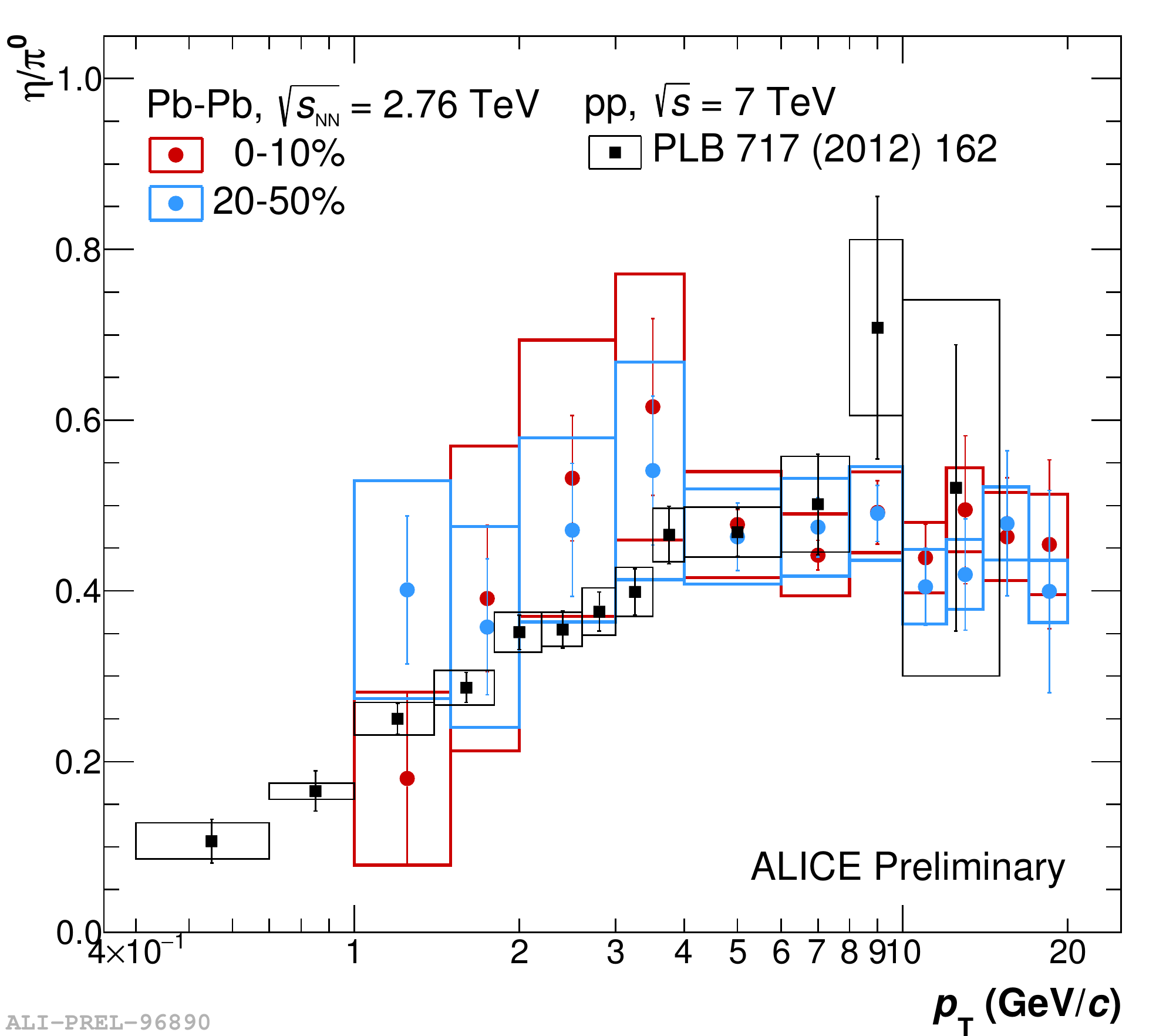}
\end{subfigure}\begin{subfigure}{0.35\textwidth}\includegraphics[width=5.0cm]{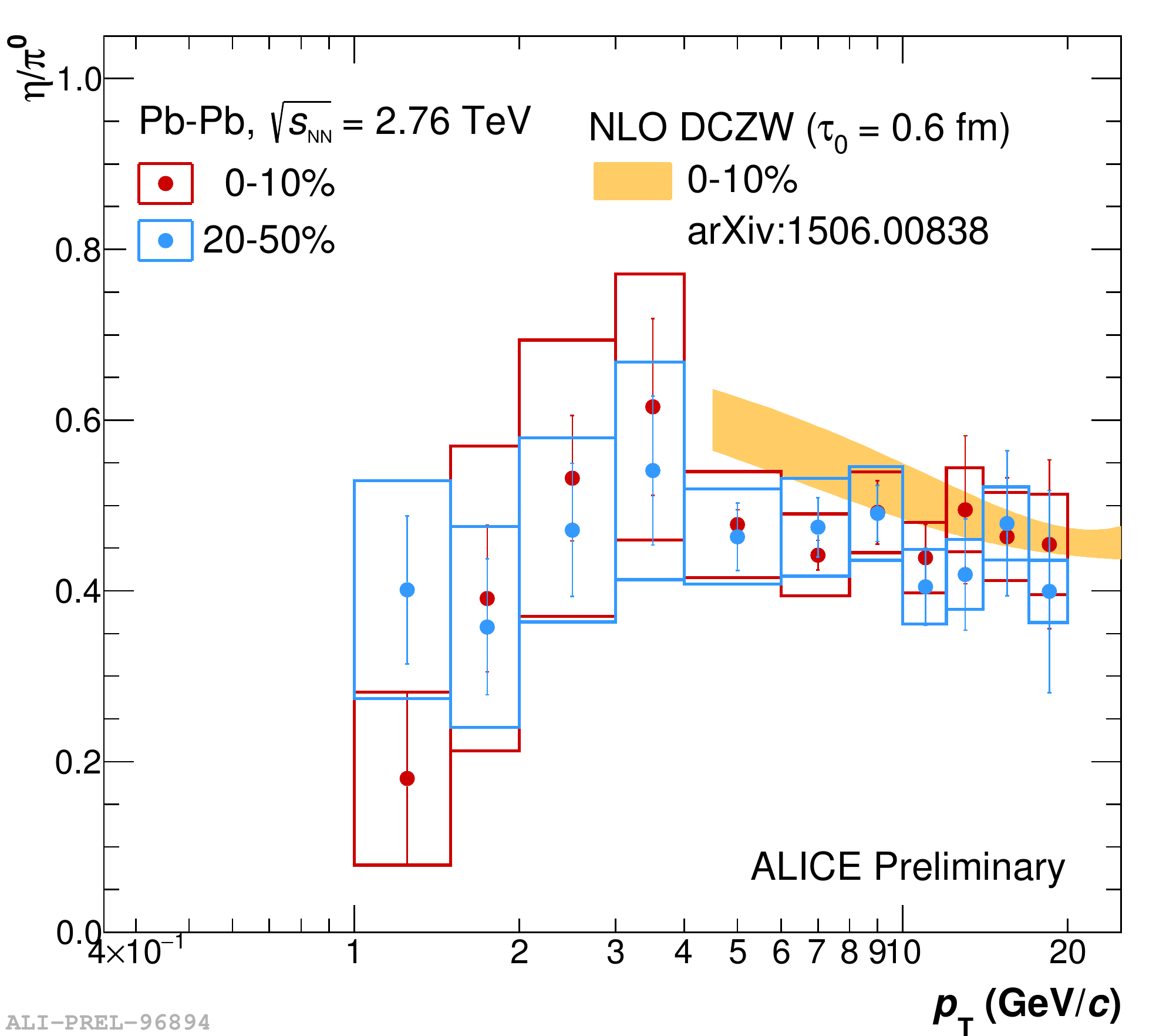}
\end{subfigure}\begin{subfigure}{0.35\textwidth}\includegraphics[width=5.0cm]{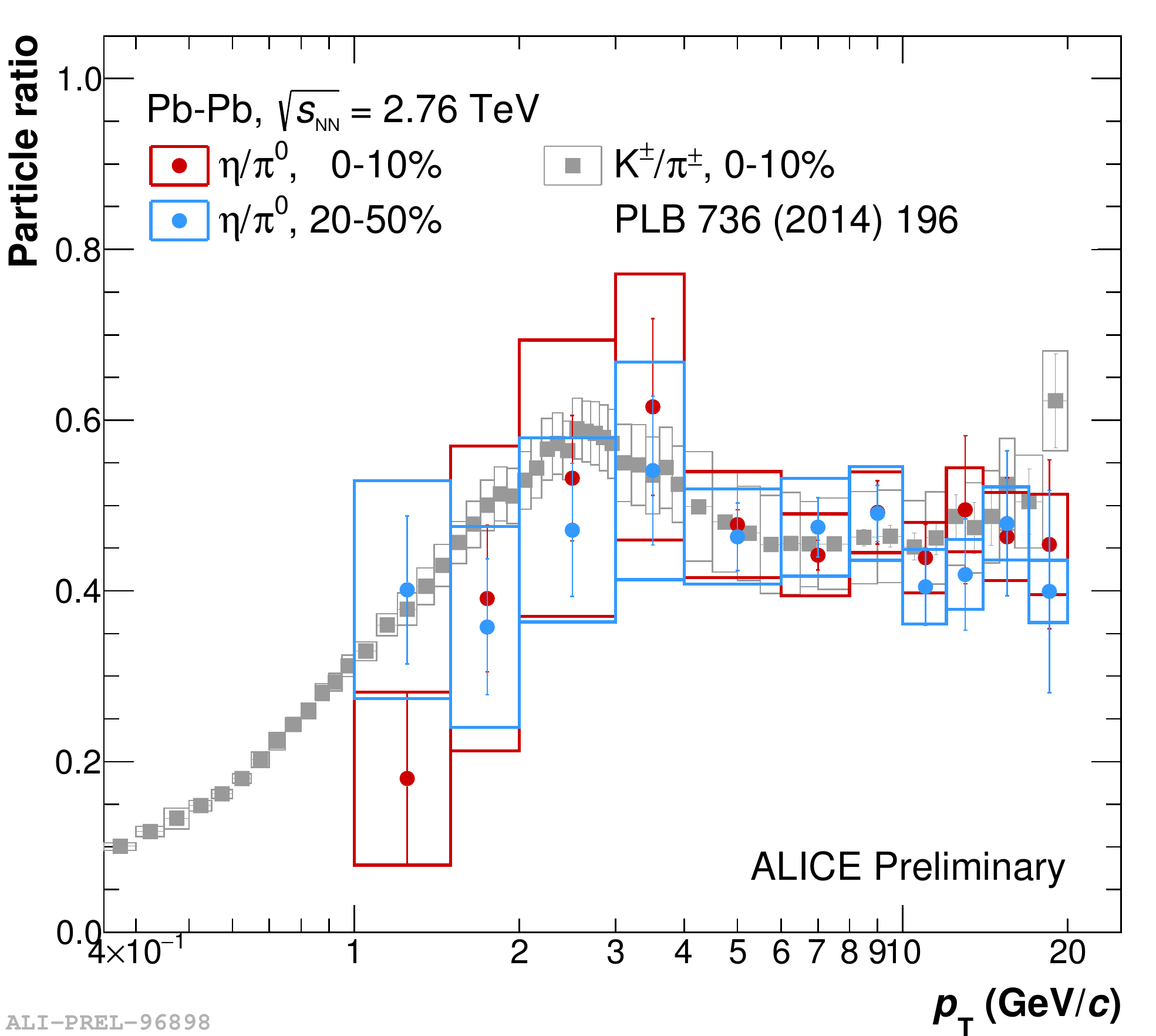}
\end{subfigure}
\caption{\footnotesize 
 $\eta/\piz$ in two centrality classes compared to measurement in pp collisions at ${\sqrt{s}} =$ 7 TeV. 
 Jet quenching predictions by Wei Dai et al~\cite{Dai:2015dxa} describe the ratio within the current uncertainties.
The trend and magnitude of the  $\eta/\piz$ ratio is consistent to previous $K^{\pm}/\pi^{\pm}$ ALICE measurements~\cite{Abelev:2014laa}.}\label{fig:newpbpbratios}
\end{figure}

\vspace{-0.2cm} 
New results on the $\eta$ meson are presented in Fig.~\ref{fig:newpbpbyields} (right). These measurements are the first and the only existing $\eta$ heavy ion measurements at the LHC. The new $\pi^{0}$ results in Fig.~\ref{fig:newpbpbyields} (left), the measurements presented are from a data set offering 10 times the integrated luminosity with respect to previous ALICE results\cite{Abelev:2014ypa}. This increase of luminosity allowed for an 
improved measurement that probes the \pt~region above 12~\GeVc~ and up to 20~\GeVc. 
The measurements in Fig.~\ref{fig:newpbpbyields} are compared to NLO pQCD predictions from both $\piz$ and $\eta$ production in pp collisions~\cite{deFlorian:2014xna} scaled by $N_{coll}$. These comparisons show a suppression in the Pb-Pb data with respect to scaled NLO pQCD.  
The $\eta/\pi^{0}$ ratio has been measured (Fig.~\ref{fig:newpbpbratios}) and compared to the ALICE pp measurement at 7~TeV CM energy,  to  corresponding $K^{\pm}/\pi^{\pm}$ measurements~\cite{Abelev:2014laa}, and to pQCD (jet-quenching based) calculations~\cite{Dai:2015dxa}. These results indicate a similar magnitude of the ratio in Pb-Pb collisions and in pp collisions and consistent with pQCD within the experimental uncertainties. No effects related to the strange quark content that may lead to discernible differences between the $\eta/\pi^{0}$ and $K^{\pm}/\pi^{\pm}$ are observed. 

{\vspace{0.75cm}
\begin{SCfigure}[\sidecaptionrelwidth][!hb]
\centering
\includegraphics[width=0.38\textwidth]{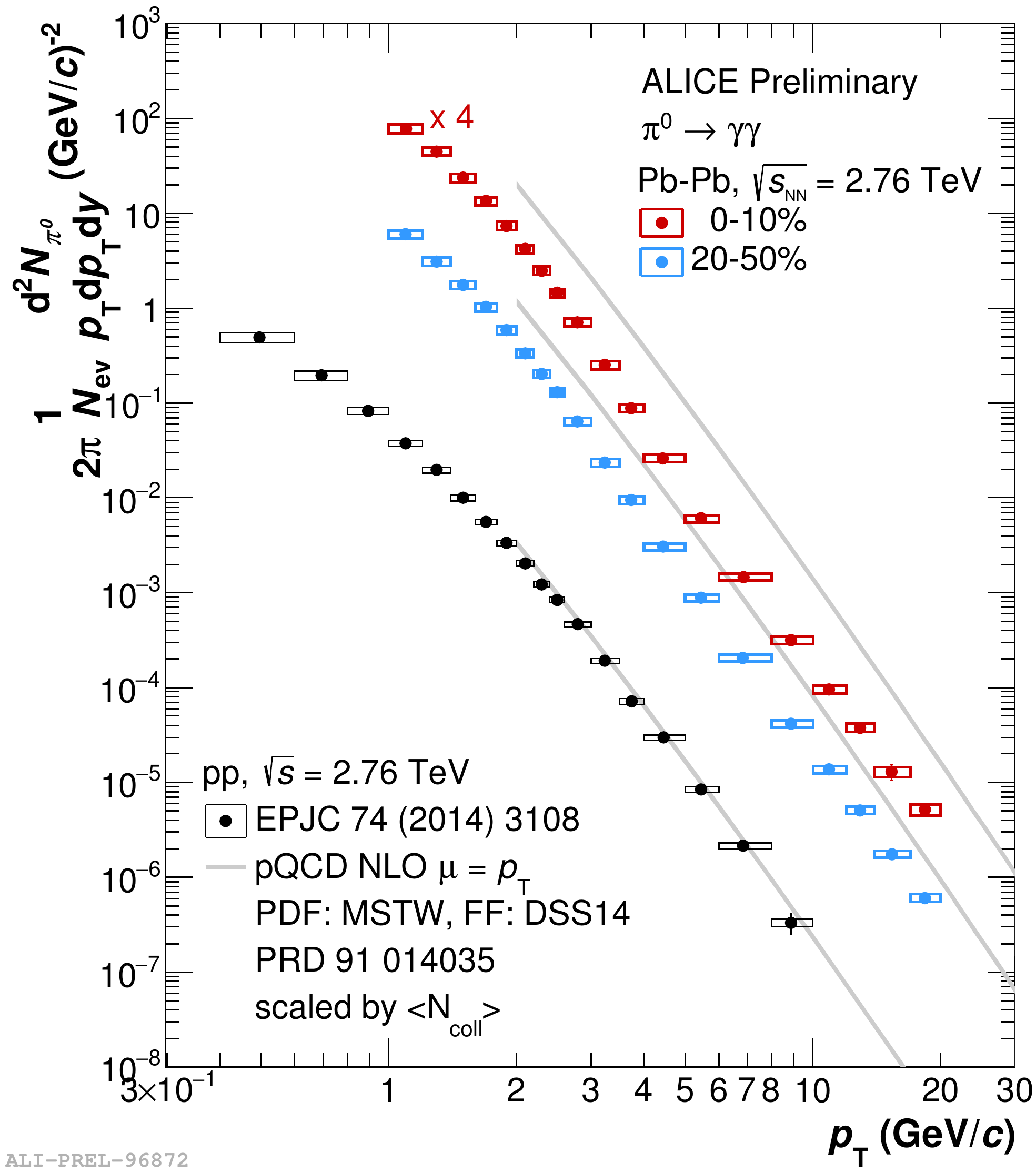}\includegraphics[width=0.38\textwidth]{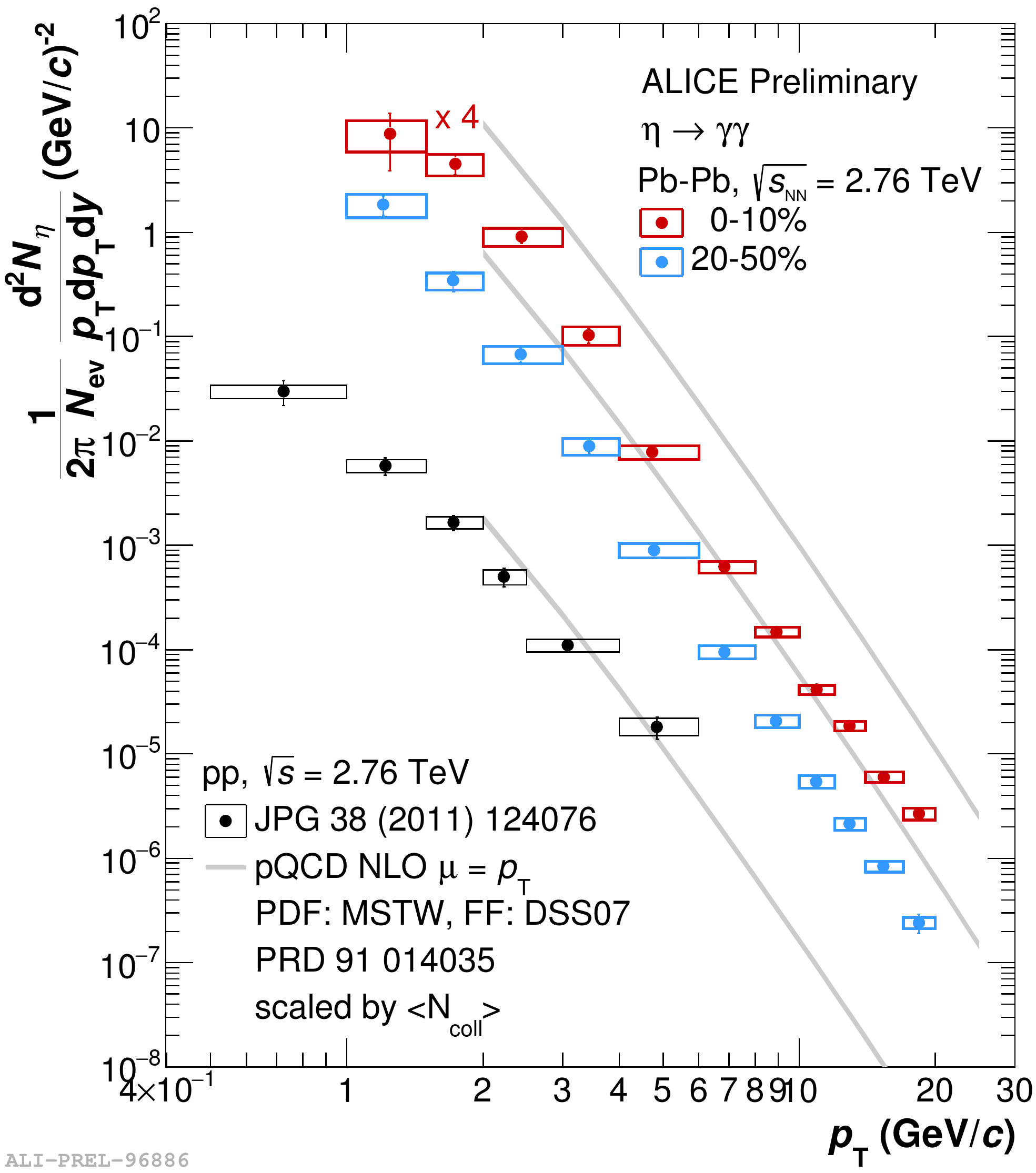}
\caption{\footnotesize Left:~$\piz$ yields measured in two centrality classes in Pb-Pb at $\sqrt{s_{NN}}=~$2.76 TeV compared to the published pp reference at the same CM energy. Right:~$\eta$ yields measured in two centrality classes compared to the pp reference at the same CM energy.}\label{fig:newpbpbyields}
\end{SCfigure}
}
\vspace{-0.0cm} 
\section{Summary}
\vspace{-0.3cm} 

ALICE measures neutral mesons in a wide \pt~range thanks to complementary detectors.
 ${\piz}$ and $\eta$ invariant yields have been measured by ALICE in \pp and \pbpb collisions.
 While the magnitude is well described by NLO pQCD calculations, there is a growing discrepancy as a function of \pt~and with increasing ${\sqrt{s}}$. A \pt~dependent suppression of ${\piz}$ is observed via the ${R_{\aa}}$ measurement. While the shape of ${R_{\aa}}$ is comparable between RHIC and LHC energies, 
at LHC energies a stronger suppression is observed indicating an energy dependence of the suppression.
 We have presented a new, and first at the LHC, $\eta$ meson measurent and an updated  $\piz$ result in Pb-Pb HIC which extend the \pt~range from \pt~$=$~12 GeV/$c$ to 20~GeV/$c$ with respect to previous ALICE measurements.
 $\eta/\piz$~ reaches a constant value for \pt~$>$4 GeV/$c$. For \pt~below 4~GeV/$c$, a complex dependence is seen, that is consistent with the one observed for $K^{\pm}/\pi^{\pm}$. No significant differences are seen between \pbpb and pp.

\renewcommand{\topfraction}{0.85}
\renewcommand{\bottomfraction}{0.85}
\renewcommand{\textfraction}{0.15}
\renewcommand{\floatpagefraction}{0.8}
\renewcommand{\textfraction}{0.1}
\setlength{\floatsep}{5pt plus 2pt minus 3pt}
\setlength{\textfloatsep}{5pt plus 2pt minus 3pt}
\setlength{\intextsep}{5pt plus 2pt minus 2pt}

\vspace{-0.5cm} 
\bibliographystyle{elsarticle-num}
\bibliography{bibliography}

\begin{thebibliography}{10}
\expandafter\ifx\csname url\endcsname\relax
  \def\url#1{\texttt{#1}}\fi
\expandafter\ifx\csname urlprefix\endcsname\relax\def\urlprefix{URL }\fi
\expandafter\ifx\csname href\endcsname\relax
  \def\href#1#2{#2} \def\path#1{#1}\fi

\bibitem{Lappi:2013zma}
T.~Lappi, H.~Mäntysaari, {Single inclusive particle production at high energy
  from HERA data to proton-nucleus collisions}, Phys. Rev. D88 (2013) 114020.

\bibitem{Dai:2015dxa}
W.~Dai, X.-F. Chen, B.-W. Zhang, E.~Wang, {$\eta$ meson production of
  high-energy nuclear collisions at NLO}, Phys. Lett. B750 (2015) 390--395.

\bibitem{Guo:2000nz}
X.-f. Guo, X.-N. Wang, {Multiple scattering, parton energy loss and modified
  fragmentation functions in deeply inelastic e A scattering}, Phys. Rev. Lett.
  85 (2000) 3591--3594.

\bibitem{ALICE}
K.~Aamodt, et~al., {The ALICE experiment at the CERN LHC}, J. Instrum. 3 (2008)
  S08002.
\newblock \href {http://dx.doi.org/10.1088/1748-0221/3/08/S08002}
  {\path{doi:10.1088/1748-0221/3/08/S08002}}.

\bibitem{PHOS}
Photon spectrometer phos, technical design report, CERN/LHCC 99-4.

\bibitem{Abelev:2014ffa}
B.~B. Abelev, et~al., {Performance of the ALICE Experiment at the CERN LHC},
  Int. J. Mod. Phys. A29 (2014) 1430044.

\bibitem{ALICE_ITS}
K.~Aamodt, et~al., {Alignment of the ALICE Inner Tracking System with
  cosmic-ray tracks}, J. Instrum.

\bibitem{ALICE_TPC}
J.~Alme, et~al., The alice tpc, a large 3-dimensional tracking device with fast
  readout for ultra-high multiplicity events, Nucl. Instrum. Meth. A622 (2010)
  316--367.

\bibitem{Cortese:2008zza}
P.~Cortese, et~al., Alice electromagnetic calorimeter technical design report,
  CERN-LHCC-2008-014.

\bibitem{Abelev:2012cn}
B.~Abelev, et~al., {Neutral pion and $\eta$ meson production in proton-proton
  collisions at $\sqrt{s}=0.9$ TeV and $\sqrt{s}=7$ TeV}, Phys. Lett. B717
  (2012) 162--172.

\bibitem{Adare:2013esx}
A.~Adare, et~al., {Spectra and ratios of identified particles in Au+Au and
  $d$+Au collisions at $\sqrt{s_{NN}}=$200 GeV}, Phys.Rev. C88~(2)  024906.

\bibitem{deFlorian:2014xna}
D.~de~Florian, R.~Sassot, M.~Epele, R.~J. Hernandez-Pinto, M.~Stratmann,
  {Parton-to-Pion Fragmentation Reloaded}, Phys. Rev. D91~(1) (2015) 014035.

\bibitem{Abelev:2014ypa}
B.~B. Abelev, et~al., {Neutral pion production at midrapidity in pp and Pb-Pb
  collisions at $\sqrt{s_{NN}}=$2.76 TeV}, Eur. Phys. J. C74~(10) (2014) 3108.

\bibitem{Glauber}
M.~L. Miller, K.~Reygers, S.~J. Sanders, P.~Steinberg, Glauber modeling in high
  energy nuclear collisions, Ann.Rev.Nucl.Part.Sci. 57 (2007) 205--243.

\bibitem{Abelev:2012hxa}
B.~Abelev, et~al., {Centrality Dependence of Charged Particle Production at
  Large Transverse Momentum in Pb--Pb Collisions at $\sqrt{s_{{NN}}} = 2.76$
  TeV}, Phys. Lett. B720 (2013) 52--62.

\bibitem{PhysRevLett.109.152301}
A.~Adare, et~al., Evolution of ${\ensuremath{\pi}}^{0}$ suppression in
  $\mathrm{Au}+\mathrm{Au}$ collisions from $\sqrt{s_{NN}} =$39 to 200~{GeV},
  Phys. Rev. Lett. 109 (2012) 152301.

\bibitem{Adare:2008qa}
A.~Adare, et~al., {Suppression pattern of neutral pions at high transverse
  momentum in Au + Au collisions at $\sqrt{s_{NN}} =$200~{GeV} and constraints
  on medium transport coefficients}, Phys. Rev. Lett. 101 (2008) 232301.

\bibitem{Adare:2012uk}
A.~Adare, et~al., {Evolution of $\pi^0$ suppression in Au+Au collisions from
  $\sqrt{s_{NN}} = 39$ to 200~{GeV}}, Phys. Rev. Lett. 109 (2012) 152301.

\bibitem{PhysRevLett.100.242301}
M.~M. Aggarwal, et~al., {Suppression of {H}igh-$p_{\rm T}$ {N}eutral {P}ion
  {P}roduction in {C}entral ${\rm Pb}+{\rm Pb}$ {C}ollisions at
  $\sqrt{s_{NN}}=17.3$ {GeV} {R}elative to $p+{\rm C}$ and $p+{\rm Pb}$
  {C}ollisions}, Phys. Rev. Lett. 100 (2008) 242301.

\bibitem{Abelev:2014laa}
B.~B. Abelev, et~al., {Production of charged pions, kaons and protons at large
  transverse momenta in pp and Pb-Pb collisions at $\sqrt{s_{NN}} =$2.76 TeV},
  Phys. Lett. B736 (2014) 196--207.

\end{thebibliography}
\vspace{-0.5cm} 






\end{document}